TABLE OF CONTENTS (TOC)

**Successful Entrapment of Carbon Dots within Flexible Free-Standing Transparent Mesoporous Organic-Inorganic Silica Hybrid Films for Photonic Applications**

Anastasia Vassilakopoulou*, Vasilios Georgakilas, Nikolaos Vainos and Ioannis Koutselas

University of Patras, Greece

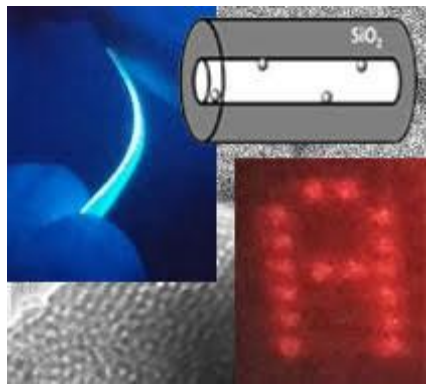

Highly luminescent carbon dots are embedded in flexible organic-silica mesoporous films, protected from high temperatures. Effortless holographic mask replication with 2% shrinkage is demonstrated

# Successful Entrapment of Carbon Dots within Flexible Free-Standing Transparent Mesoporous Organic-Inorganic Silica Hybrid Films for Photonic Applications

Anastasia Vassilakopoulou*[✉], Vasilios Georgakilas, Nikolaos Vainos and Ioannis Koutselas

*Materials Science Department, School of Natural Sciences, University of Patras, Patras, 26504, Greece*

[✉]*Anastasia.vassil@gmail.com*

## ABSTRACT

The effective entrapment of Carbon dots (CDs) into a polymer-silica hybrid matrix, formed as free standing transparent flexible films, is presented. CDs of 3 nm mean size with strong photoluminescence are embedded into a silica matrix during the sol-gel procedure, using tetraethylorthosilicate and F127 triblock copolymer as the structure directing agent under acidic conditions. The final hybrid nanostructure forms free standing transparent films that show high flexibility and long term stable CDs luminescence indicating the protective character of the hybrid matrix. It is crucial that the photoluminescence of the hybrid's CDs is not seriously affected after thermal treatment at 550 $^0$C for 30min. Moreover, the herein reported hybrid is demonstrated to be suitable for the fabrication of advanced photonic structures using soft lithography process due to its low shrinkage and distortion upon drying.

**Introduction**

Carbon dots (CDs) is a new class of nanostructured materials that have recently attracted general interest for their unique properties of strong tunable photoluminescence, low cost, low toxicity and biocompatibility [1]. These properties have led to a series of potential applications as in light emitting diodes [2,3], solar cells [4,5], sensing [6], catalysis [7], integration in photovoltaic devices etc. [8] and more importantly to a possible breakthrough in biosensing, bioimaging [9- 13] and medical diagnosis [14].

Usually CDs are quasi-spherical nanostructures 2-10 nm, consisted from amorphous carbon core that contains well embedded smaller graphitic and turbostratic parts. The external surface is covered mainly by carboxylates and several others organic functional groups such as hydroxyl, amines and amides depending on the preparation methods and the precursor molecules. The presence of hydrophilic groups at the surface of CDs induces excellent solubility in water. The main characteristic of CDs is the relatively strong photoluminescence, which is mainly depends on their size, the excitation wavelength and the surface functionalization [6, 15].

Incorporation of highly luminescent C dots in appropriate solid state matrices is highly attractive in order to materialize their properties in real device applications. Stabilization in solid materials such as polymers, ceramics, and inorganic oxides is an important challenge since often fluorescence quenching pathways are presented due to CDs aggregation or other chemical interactions of its external surface. In the literature, there are several examples where CDs were successfully incorporated into polymer matrices such as polymethylmethacrylate (PMMA) [16], polydimethylsiloxane (PDMS) [17], polyvinyl pyrrolidone (PVP) [18], polyvinyl alcohol (PVA) [12,19] or gel glass [20]. In most cases the embedded CDs in solid polymeric films showed stable photoluminescence. Recently, Rogach et al [18] presented the formation of a white light-emitting diode with excellent color stability, combining blue-emitting CDs with green and red-emitting polymer dots in polyvinyl pyrrolidone (PVP) matrix. Among the candidate solid matrices, silica gel is particularly attractive due to its optical properties and inherent stability which would allow direct incorporation into existing lighting formats [21].

In the present article, the incorporation of CDs in a polymer/silica hybrid layered material is presented for the first time. The CDs were added in the reaction solution during the sol gel formation of the polymer/silica hybrid and entrapped in the interface between the two components and in the silica part. The effective entrapment resulted in the formation of a free standing transparent flexible film that emits strong fluorescence by excitation. The presented hybrid film protects the CDs even when treated at high temperatures, such as 550$^o$C, as evidenced from their strong PL. Finally, the hybrid films can successfully replicate holographic surface relief structures at the micro/nano level, while retaining the optical properties of the embedded CDs, making them suitable for complex photonic device fabrication. The soft lithographic process is performed with no applied pressure and shows low shrinkage upon drying, due to its porous framework. As it is known from the literature, in order for a material to be compatible with soft lithography it needs to be able to completely fill a pre-patterned stamp, usually fabricated on a silicon wafer and maintain its volume during curing or drying yielding smooth structures, accurately replicating the surface relief structure of the master stamp [22].

1. **Experimental**

**1.1. Materials**

All chemicals were purchased from Sigma-Aldrich. Nonionic block copolymer EO97PO67EO92 (F127, Mw=12600), Tetraethyl orthosilicate (TEOS, Mw=208.33), n-hexadecyltrimethylammonium bromide (CTAB, MW=364.45), aqueous ammonia solution wt. 32%, hydrochloric acid (HCl), absolute ethanol (Eth, 99%), citric acid (CA, Mw=192.12), triethylene tetramine (TETA, Mw=146.23). All chemicals were used without further purification.

**1.2 CDs synthesis**

350 mg of citric acid and 350 mg of triethylene tetramine were diluted in 10ml of deionized water and the solution was heated in a 600 W commercial off the shelf microwave oven for 3 minutes. After cooling the final solution was purified by dialysis using a membrane with 2000 MWCO for 5 days.

**1.3 Preparation of transparent flexible films within Carbon Dots (Silica/F127-CDs) and mesoporous silica with embedded CDs (Silica-CDs):**

Transparent flexible films with CDs (Silica/F127-CDs) were prepared by the standard sol-gel technique using TEOS as a silica precursor and F127 triblock copolymer as the structure directing agent. In a typical synthesis, 2 g of F127 was dissolved in 5 ml of absolute ethanol with 0.15 g of 2M HCl solution. Afterwards, 1ml of aqueous solution of CDs and finally 6 ml of TEOS was added slowly in the reaction mixture under continuous stirring. The sol was transferred in a vacuum chamber for 10 minutes and then was kept at room temperature for 2 days for aging on polystyrene dishes. The final product has the form of a transparent flexible film which was easily lifted off the substrate. Spin coating the sol solution on quartz glass at 3000 rpm for 20 seconds led to thin films of uniform micrometric thickness as deduced from optical microscopy. The Silica/F127-CDs composite was then heated at 550$^o$C where the organic polymer was removed affording a porous silica matrix with CDs embedded (Silica-CDs). For comparison, a transparent flexible film based on an organic/inorganic hybrid with silica and F127 polymer, without CDs was also prepared using the same sol gel procedure and quantities as above (Silica/F127).

For the soft lithography process, a computer generated hologram was patterned on a silicon wafer using standard e-beam lithography and reactive ion etching process which was treated with tridecafluoro-1,1,2,2-tetrahydroctyl vapor to provide antisticking properties. Afterwards the Silica/F127 hybrid gel with or without CDs, was poured on the silicon stamp and left to dry for four hours under ambient atmosphere, without any external pressure. Then, the patterned free-standing films were easily detached using tweezers.

### 1.4 Characterization

X-ray powder diffraction (XRD) measurements were carried out on a D8 Advance system (Bruker) with Cu ($\lambda$=1.54178 Å) radiation, without using a monochromator. The voltage and current applied to the X-ray tube were set at 40kV and 40mA. The XRD patterns were measured in the 2θ range from 2º to 80º. The step rotation was 0.02º and the scanning speed was maintained at 1 sec/step. Infrared spectra were measured on a Fourier transform spectrometer (Digilab) in the form of powder milled with anhydrous KBr and pressed into pellets. The final spectra were the average of 20 scans in the frequency range of 400-4000 cm$^{-1}$, at 2 cm$^{-1}$ resolution. Optical absorption (OA) spectra in the UV-Vis spectral region were recorded on a Shimadzu 1650 spectrophotometer in the range of 200-800 nm, at a sampling step of 0.5 nm at 1.5 nm slits, using a combination of halogen and deuterium lamps as sources. Reflectance and transmission spectra were also obtained with the usage of the related integrating sphere for evaluating the transmittance. The photoluminescence (PL) spectra were obtained from solid pressed pellets, thick deposits or solutions on quartz plates and cuvettes, mounted in a Hitachi F-2500 FL spectrophotometer employing a xenon 150 W lamp and a R928 photomultiplier. The excitation and detection slits were set at 2.5 nm and the accelerating voltage was set to 700 V. The index of refraction was measured with an Abbe Kruss AR2008 refractometer.

The thermogravimetric analysis of the samples was performed with thermal analyzer TG Q500 (TA Instruments) under ambient air with step 50 $^{o}$C/min up to 800 $^{o}$C.

The morphology of the samples was observed by scanning electron microscopy (SEM) (EVO-MA10, Bruker) and transmission electron microscopy (TEM) (JEOL-JEM 2100) where pictures were obtained with camera (Gatan).

### 2. Results and Discussion

The hydrothermal heating of citric acid and triethylene tetramine led to the formation of highly water soluble CDs, with a mean diameter of 3 nm as estimated from TEM images (see Figure 2, right). The CDs solution exhibits strong PL with maximum at 485 nm when excited at $\lambda_{ex}$=350nm (see Figure 5). The main step of the procedure was a sol gel technique where CDs were added to the ethanolic solution of F127 block copolymer without affecting the dispersibility of the polymer or the micelles formed. Finally, after the addition of TEOS –the silica precursor- and the related hydrolysis step, CDs remained homogenously dispersed in the reaction mixture. The final mixture was spread on a polystyrene surface and air-dried. This procedure provided a light yellow, highly transparent and flexible hybrid film (Silica/F127-CDs). The Silica/F127-CDs hybrid was then heated at 550$^o$C where the organic material -F127 block copolymer- was removed affording a mesoporous silica matrix with CDs embedded (Silica-CDs) as deduced from the active photoluminescence signal. The procedure and the resultant material can be schematically depicted in Figure 1.

Figure 2 (left) shows the powder XRD patterns of (a) Silica/F127-CDs, and (b) Silica-CDs. The broad peak centered at 23.81$^o$ for the Silica/F127-CDs is attributed to highly disordered carbon atoms [12]. The peak position reflects an amorphous framework in the CDs, however, indicating order having a lattice plane spacing of 0.37nm which corresponds to the graphite (002) planes [23,24]. The pristine graphite (002) plane spacing is about 0.33nm (26.64$^o$), thus, the slight increase detected can be explained as arising from a possible poor crystallization [25].

After the heat treatment (Silica-CDs) the peak is shifted to 30$^o$ and slightly increased in width. The latter can be understood to occur as the structure of CDs are broken down during the thermal treatment as well as due to the removal of the carbon dots protective layers. Both these effects lead to smaller structures and increase the number of stacking faults, thus, leading to wider observed XRD peaks. The shift of the XRD peak to 30$^o$ is probably due to a phase change of the remaining CDs where the (002) planes can collapse closer due to the thermally induced degradation. Broad XRD peaks covering the 30$^o$ have also been observed elsewhere [26]. The broad shoulder appearing at c.a. 40$^o$ may be attributed to the (10) and (101) graphitic planes of the remaining CDs [27]. In Figure 2(right), a representative image is shown of the CDs as obtained from solution, where the diameter is calculated to be 3nm.

FT-IR spectroscopic data, presented in Figure 3, are provided for the (a) CDs, (b) Silica/F127-CDs and (c) Silica-CDs. The spectrum of CDs in line (a) has mainly three characteristic peaks; two strong peaks being at 1550 cm$^{-1}$ (N-H) and 1660 cm$^{-1}$ (C=O stretching) that were attributed to the amide bond which is often appearing at the protective layer of several CDs and a broad band centered at 3430 cm$^{-1}$ assigned to stretching vibration of N-H, with a small shoulder at 3270 cm$^{-1}$ attributed to OH stretching vibration. Line (c) shows the FT-IR spectrum of the silica-CDs hybrid film which was formed after the thermal treatment of Silica/F127-CDs and the removal of F127 polymer. Silica-CDs hybrid consisted of a SiO$_2$ skeleton and embedded CDs. The characteristic strong peaks at 1020 cm$^{-1}$, 950 cm$^{-1}$ and a smaller one at 790 cm$^{-1}$ were attributed to the Si-O-Si bridge vibrations originating from the silicate network. The remaining peaks that are due to the presence of CDs have shifted to 1640 and 1730 cm$^{-1}$ and are attributed to the C=O vibrations. This demonstrates that the amide groups at the surface of the CDs have been transformed to ketones and carboxylates after the heat treatment. After the elimination of the amides, the broad peak was shifted to lower wave numbers and centered at 3350 cm$^{-1}$ (OH stretching vibration). Line (b) shows the FT-IR spectrum of Silica/F127-CDs hybrid. The peaks centered at 2880 cm$^{-1}$, 1350 cm$^{-1}$ and 1300 cm$^{-1}$ (stretching and bending vibrations of C–H bonds) and the broad band centered at 3270 cm$^{-1}$ (OH stretching vibration) were attributed to the directing agent, F127 molecule, which is a dominant species in the hybrid. The remaining peaks show the existence of the silicate network similarly to line (c).

The optical properties of Silica/F127-CDs were studied by means of UV-Vis (Figure 4) and photoluminescence (PL) spectroscopies (Figure 5). As demonstrated in Figure 4(line b), Silica/F127-CDs showed strong optical absorption (OA) in the UV region with two absorption bands centered at 241 nm (π-π* transition) and 363 nm (n-π* transition) [28], with full widths at half maximum (FWHM) of 79.9 and 32.5 nm, respectively, where in the reference spectrum of pristine mesoporous material no such peaks are observed in the whole UV-Vis range (see Figure 4 line a). The absorption spectrum, of the heat treated Silica/CDs (see Figure 4 line c) shows a broad featureless absorption profile, however, with some minor characteristic features at 300nm that have been blue shifted from the 363nm peak. This is in accordance with the elimination of the organic groups that is also indicated by the corresponding FTIR spectrum.

Figure 5 shows the PL characteristics of Silica/F127-CDs (a) and silica-CDs hybrids (b). Silica/F127-CDs exhibited a strong blue PL response, with peak centered at 466 nm when the excitation wavelength was 350 nm (see Figure 5a). It should be noted that there no excitation dependence PL behavior was observed for the pristine water CDs solution as well as with the transparent flexible Silica/F127-CDs film.

Both absorption and PL peaks in the spectra of Silica/F127-CDs hybrid are due to the existence of CDs, since the starting porous matrix of pure hybrid silica-F127 material exhibits no absorption peaks nor any PL peaks for excitation wavelengths of 350nm. Comparing the emission bands of free CDs in solution with that embedded in the silica/polymer matrix in the Silica/F127-CDs hybrid, a blue shift was observed from 483 nm to 466 nm. In a blank experiment, adding F127 co-polymer ethanol solution into the CDs water solution, a similar trend was also recorded, revealing that non covalent interaction of F127 co-polymer with the external groups of CDs is responsible for the observed blue shift and an increase of the QY of CDs luminescence. It is important that ethanol addition has no effect on the PL peak. It is often observed that there is a small shift of the PL maxima peak position with respect to the varying environment of the CDs which may include pH, solvent and/or other organic molecules [15, 29, 30]. The quantum yield (QY) of the CDs in solution was about 40% while embedded in the Silica/F127 matrix the QY is slightly increased to 50 %. This notable increase is attributed to the interaction of CDs with the F127 co polymer in the solid matrix as revealed by the previous blank experiment that described above.

The Silica-CDs film, that were derived after the thermal treatment of Silica/F127-CDs hybrid and the removal of F127 has still a strong blue, naked eye visible, PL emission with a maximum shifted at 426 nm, under the same excitation wavelength at $\lambda_{exc}$=350 nm (see Figure 5 line b). The thermal heating of Silica/F127-CDs hybrid apart from the removal of F127 copolymer was responsible for the elimination of several groups from the surface of CDs as indicated from their FTIR spectrum (see

Figure 3 line c). This structural modification resulted in the blue shifting of the PL emission maximum from 466 nm to 426 nm. This 40nm shift is most probably attributed to the chemical changes as the outer part of the CDs is being removed. Moreover, the nature of the chemical bond breakup that is induced by the heat treatment also changes the external surface chemistry of CDs (see FTIR spectrum in Figure 3c), thus, forming new chemical bonds and hybridizations which lead to the higher energy PL emission. The surface chemistry before the heat treatment has $NH_2$ and amide groups, while after the heat treatment F127 and the organic layer around C Dots were removed leaving only ketones and carboxylates. PLE spectrum for the Silica-CDs sample, not included, shows only a small peak centered at 370nm, in accordance to the high energy peak of the spectrum Fig.5d, which is probably responsible for the sole emission at 426nm of the Silica-CDs sample.

Finally, in the inset of Figure 5, the PL spectrum of Silica/F127-CDs is shown as obtained at 77K. It appears that the 77K dynamics induce a small PL shift to higher energies, of c.a.5nm, possibly by inducing phase changes to the CDs by decreasing the bond distances, thus, enhancing the band to band excitation energy. Similarly, the 77K spectra of Silica-CDs samples do not differ from those acquired at room temperature in terms of peak position, while the spectra recorded at 77K is more intense. The PLE spectrum, equivalent in principle to that of the absorbance, of Silica/F127-CDs exhibits peak at 360nm, in unison with that of the absorption spectrum, however, it does show a broad peak centered at 410nm (where the emission wavelength is set to $\lambda_{em}$=520nm) overlapping with the peak at 360nm. This broad peak reveals that there are strongly emitting states positioned at the tail of the absorption spectrum beyond 430nm, which is the absorption spectrum defined energy gap. It is probable that these states, not easily detectable in the absorption spectra, play an important role in the luminescence process of the here reported CDs.

The analysis curves of TGA for Silica-F127 hybrid film (a) and Silica/F127-CDs (b) are shown in Figure 6, normalized to the specimen's weight at 100°C. Specimens lost a small portion of their weight under 100°C that was due to the removal of trace water and ethanol (not shown). TGA analysis shows that the CDs loaded Silica/F127-CDs, (see Figure 6 line a), has a slight different behavior from that of the pristine material Silica-F127 hybrid film, (see Figure 6 line b). The TGA analysis suggested that both samples lose weight after 170°C but at slight different rate, the Silica/F127-CDs being the one losing more weight which is attributed to the partial decomposition of the external surface of the CDs. Both films continue loosing material as temperature increases but with Silica/F127-CDs being more resistant to loss. At the temperature range of 200-450°C both samples loose part of the F127 and probably part of the capping molecules of the loaded CDs and possibly some CDs. Due to the complexity of the mixture it is difficult to separate these processes, thus, the two sample show different rates of weight loss, but always in a complementary fashion. At temperatures beyond 350°C up to 450°C, the Silica/F127-CDs composite shows a faster rate of material loss due to the degradation part of the CDs. At temperatures above 450°C the final organic components are being degraded, however, the Silica/F127-CDs shows some extra weight which is, as evidenced from the PL spectra, attributed mainly to the CDs that have remain unaffected by the thermal treatment.

The flexible and transparent nature of the final films, as well as its strong PL emission under UV excitation, can be observed in the images provided in Figure 7. The flexibility of the films remains even after storing the films for six months under standard laboratory conditions under a plastic dust cover.

The mesoporous structure of the Silica/F127-CDs and Silica-CDs are shown in TEM images presented in Figure 8(a) and 8(b), respectively. The pore diameter in Silica/F127-CDs was measured to 6 nm and increased to 6.2 after calcinations in Silica-CDs, while the pore distance was measured to 11 nm and decreases to 7.2 nm after calcinations. The pores belong to a 2D or 3D hexagonal cage structure. In both hybrids, the pores were larger than the average CDs size, allowing their successful encapsulation during the mesoporous synthesis as well as the retention of the CDs throughout any handling processes. The black spots in Figure 8 were probably regions of concentrated CDs within pores (a), while in (b) the CDs were smaller in size and in some cases appeared as black spots within the pores as well. It is obvious that CDs outside the pores have been completely consumed by the heat treatment at 550°C.

The hybrids under study were evaluated for micrometer range pattern replication using the soft lithography process. Material shrinkage was found to be minimal after film dry, making possible to successfully transfer the microstructure patterning of a surface relief computer generated hologram stamp etched on a silicon wafer. The silicon, which was fabricated using standard techniques, was gently poured on with a freshly deposited hybrid droplet and remained, without any applied pressure, and remained to dry for 4 hours under ambient conditions before removing the silicon stamp. The reconstruction of the hologram under laser illumination on a screen placed at 45 cm from the film is shown in Figure 9 (image (a) at λ=650 nm and (b) at λ=404 nm) while scanning electron micrograph of the patterned film is shown in Figure 9 (right).

Features of 1 μm size of the silicon master have been successfully transferred on the Silica/F127-CDs film, where the corresponding pixel was measured to have about 0.98μm top dimension average; this yields a maximum distortion of up to 2 %. In addition, the index of refraction (**n**) of the Silica/F127 is found to be 1.432, as measured using an Abbe refractometer, which is lower than that of pure silica 1.458 due to the porous nature. Similarly, **n** for the Silica/F127-CDs is found to be 1.428 which is within the experimental error, when compared to the Silica/F127. Finally, as far photonic applications are concerned, first estimate of the transparency of a 250μm thick film was of the order of 91.8% in the 550-600nm optical range.

At end, it is worthwhile to mention that the final materials as well as the procedure to fabricate the composites and the devices are non-toxic.

## 3. Conclusion

In summary, fully hydrophilic CDs can be successfully embedded in a polymer/silica mesoporous hybrid matrix using the sol gel process, resulted in a free standing transparent flexible film that emits strong fluorescence by excitation. The entrapped CDs maintain their photoluminescence properties even after the thermal treatment of the film at $550^0$C, indicating a strong protective character of the hybrid matrix. Finally, Silica/F127-CDs hybrid has been found to yield minimum shrinkage, thus, been suitable for soft lithographic micro/nanostructuring. The fabrication of an advanced photonic structure using the soft lithography process has been successfully demonstrated.

**Acknowledgments**

The authors would like to thank Dr. Miltos Vasileiadis for his help with the soft lithography experiments.

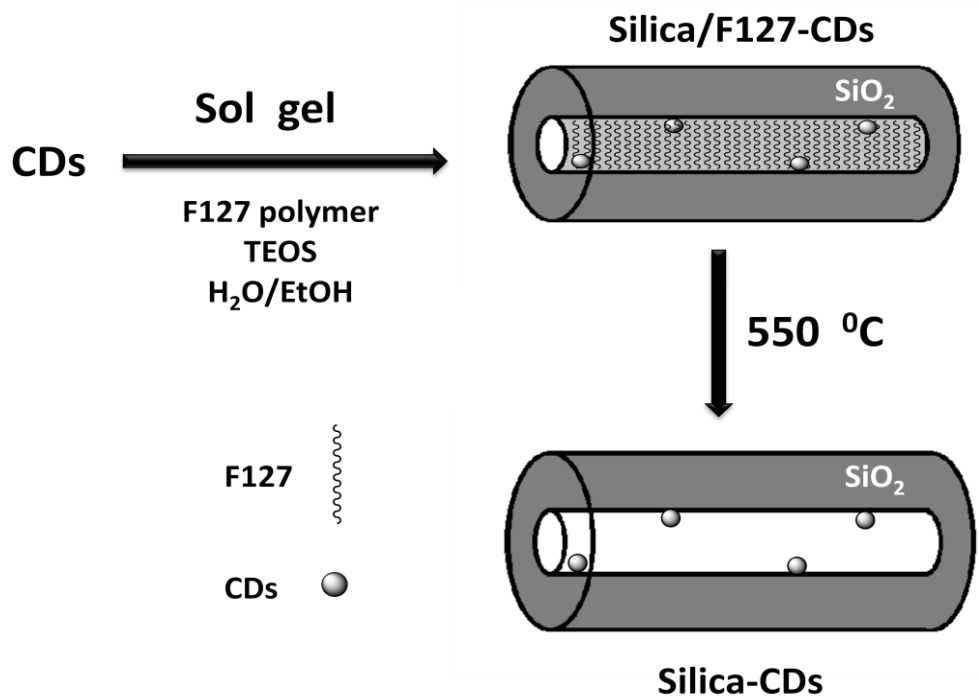

**Figure 1** Schematic representation of the preparation of Silica/F127-CDs and Silica-CDs hybrids.

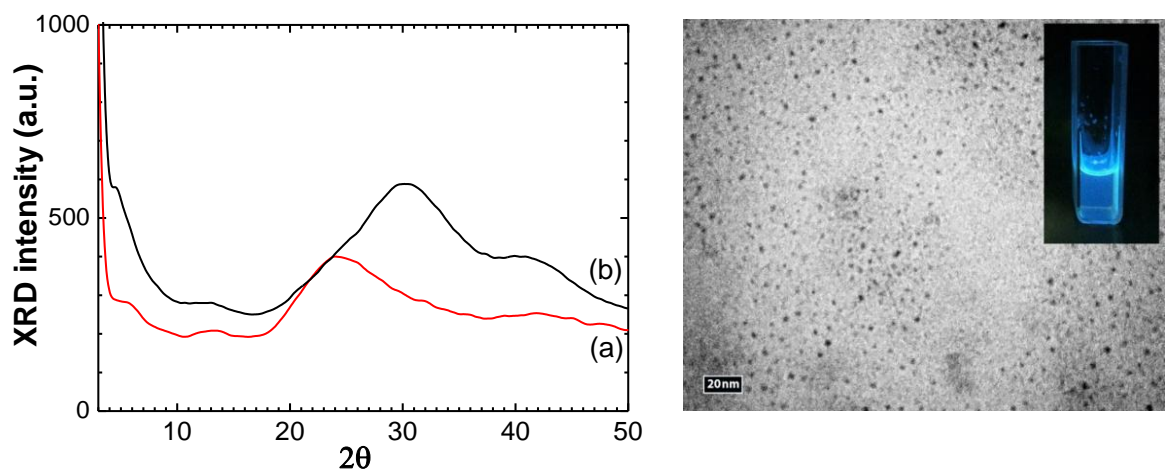

**Figure 2** (left) X-Ray powder diffraction of (a) Silica/F127-CDs and (b) Silica-CDs. (right) TEM image of the CDs from solution (bar at 20 nm). Inset shows the CDs solution excited with 365 nm.

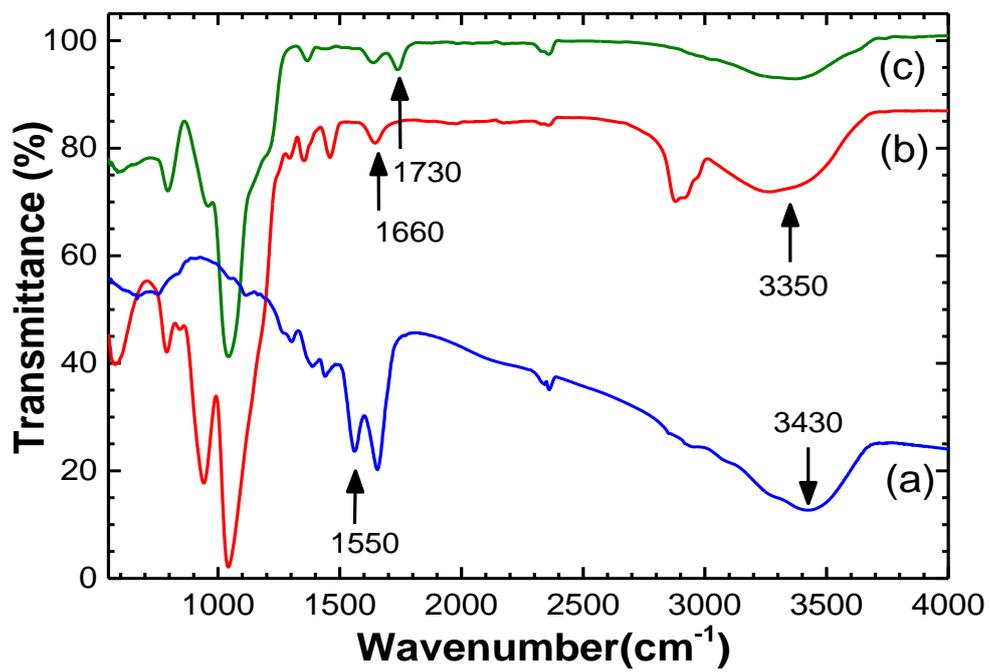

**Figure 3** FT-IR spectrum of (a) CDs, (b), Silica/F127-CDs and (c) Silica-CDs.

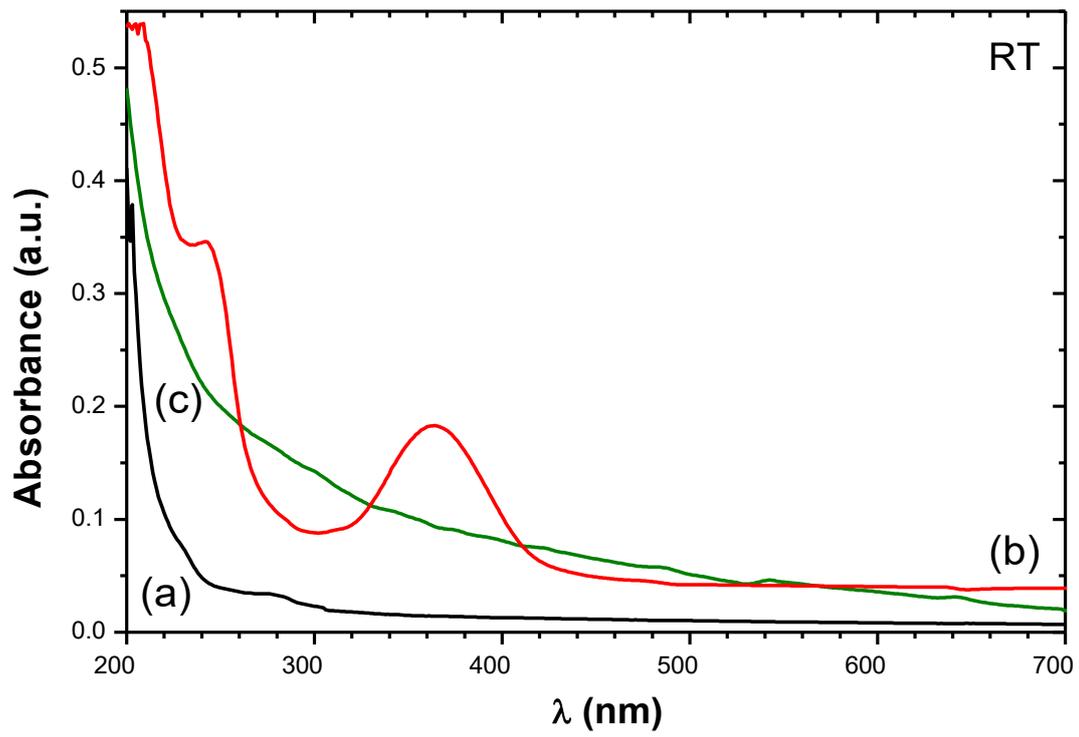

**Figure 4** UV-Vis absorption spectrum of (a) Silica/F127, (b) Silica/F127-CDs, (c) Silica/CDs.

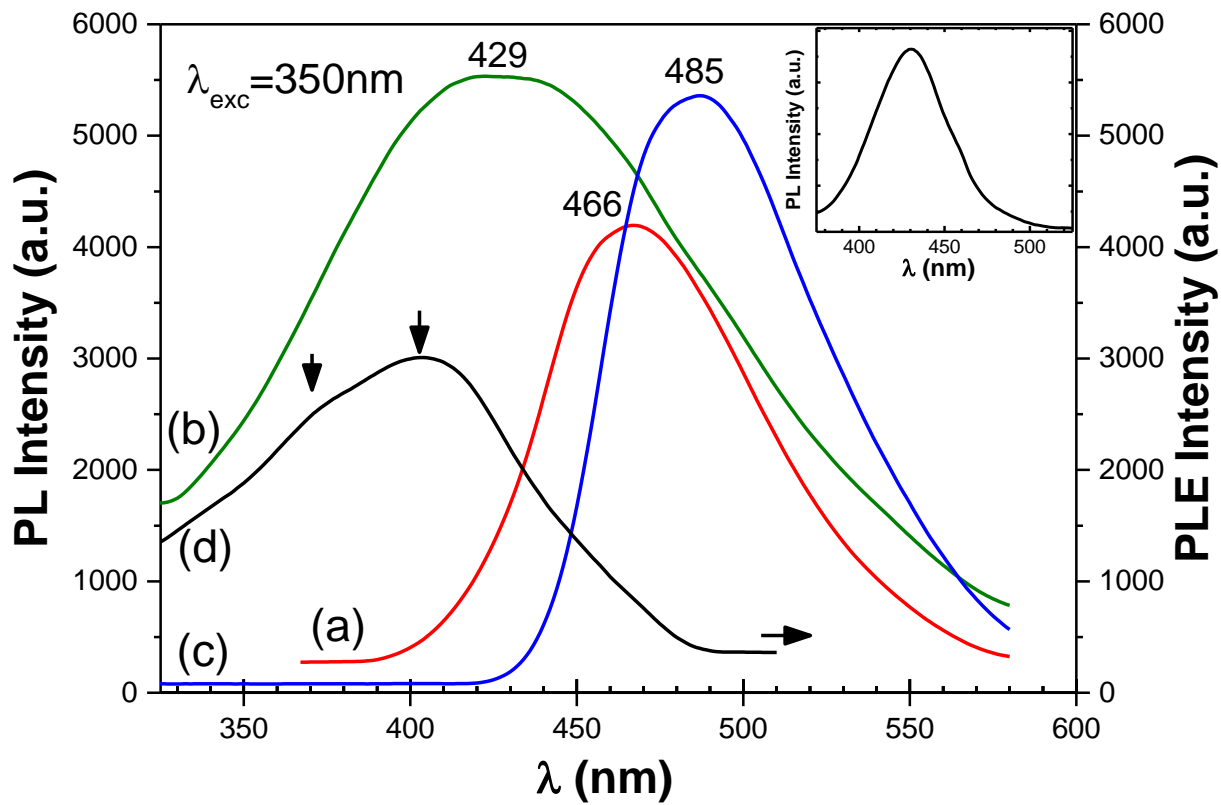

**Figure 5** PL spectra of (a) Silica/F127-CDs, (b) Silica-CDs and (c) CDs solution; (d) PLE spectrum of Silica/F127-CDs for $\lambda_{em}$=520nm. Inset shows PL spectrum of Silica/F127-CDs at 77K.

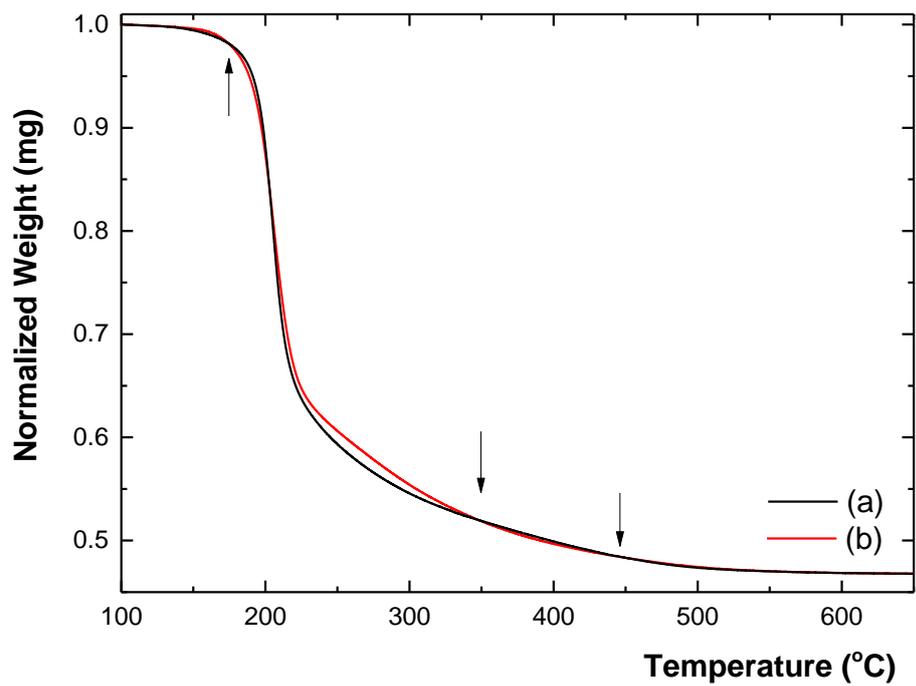

**Figure 6** TGA curves of Silica-F127 (a) and Silica/F127-CDs (b) hybrid. The TGA graphs shown have been normalized with respect to the sample value weight at 102°C, for excluding adsorbed water effects.

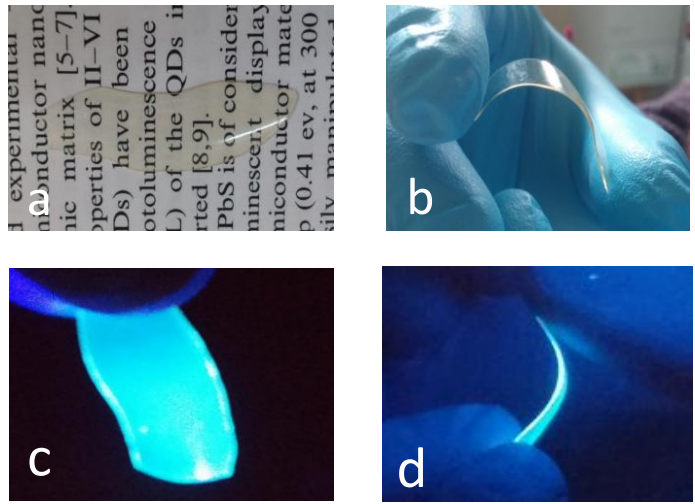

**Figure 7** Digital photographs of Silica/F127-CDs under visible (a,b) and UV(365 nm) light (c,d) showing the transparent and flexible form of the film (a,b) as well as its strong eye visible PL signal (c,d).

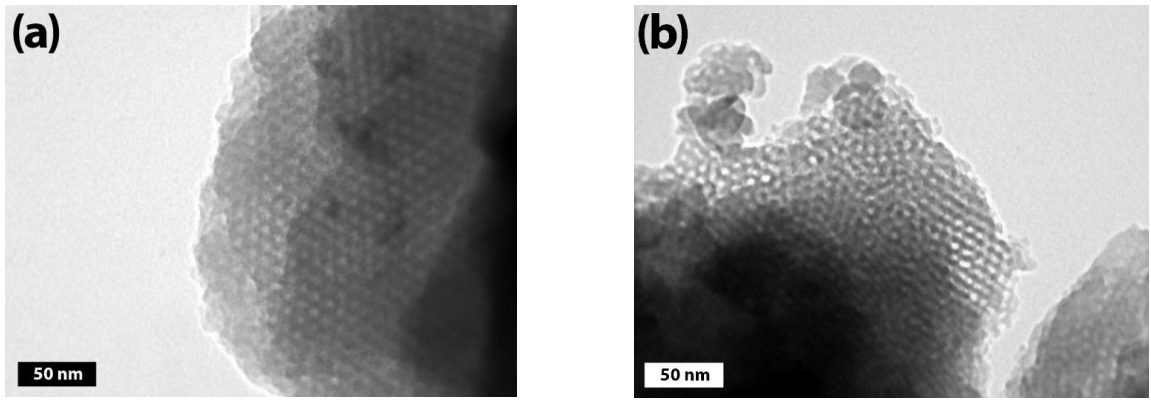

**Figure 8:** TEM micrographs for the (a) Silica/F127-CDs and (b) Silica-CDs mesoporous hybrid films.

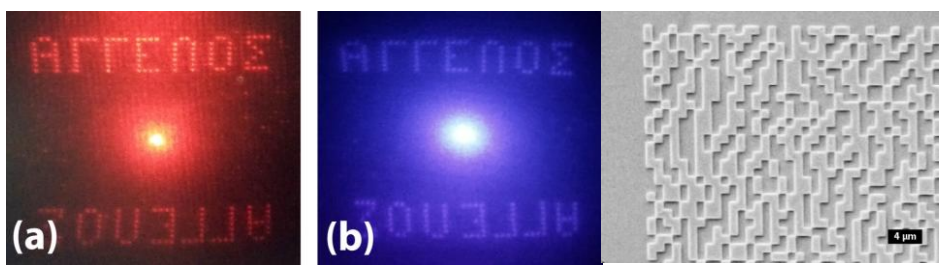

**Figure 9** (left) The reconstruction of the surface relief hologram under laser illumination at (a) 650nm and (b) 404nm. (right) Scanning electron micrographs of the patterned hybrid film.